\documentstyle[preprint,aps]{revtex}
\tightenlines
\begin{document}
\title{Electron attachment to valence-excited CO}
\author{Sanjay Kumar\footnote{Present address: Chemistry Department, Indian Institute of Technology Madras, Chennai 600~036, India}  and D. Mathur}
\address{
Tata Institute of Fundamental Research, Homi Bhabha Road,
Mumbai 400 005, India.}
\date{\today}
\maketitle
\begin{abstract}
The possibility of electron attachment to the valence $^{3}\Pi$ state of CO is examined using an {\it ab initio} bound-state multireference configuration interaction approach. The resulting  resonance has $^{4}\Sigma^{-}$ symmetry; the higher vibrational levels of this resonance state coincide with, or are nearly coincident with, levels of the parent $a^{3}\Pi$ state. Collisional relaxation to the lowest vibrational levels in hot plasma situations might yield the possibility of a long-lived CO$^-$ state. 
\end{abstract}

Negatively-charged atoms and molecules have been extensively studied over the decades \cite{massey}. However, developments of new experimental and theoretical methodologies have initiated a resurgence of interest \cite{list1} in small anionic species such as N$_2^{~-}$ and C$_2^{~2-}$ \cite{list2,list3}. We report here the results of a theoretical study of CO$^-$ anions that seeks to address two issues: (i) the long-pending characterization of a ``dipole-dominated resonance" observed 25 years ago in e-CO scattering by Wong and Schulz \cite{wong}, and (ii) rationalization of very recent observations of metastable CO$^-$ in mass spectrometric probes of transient plasmas produced in sputtering \cite{gnaser} and in interactions of molecules with intense laser fields \cite{vrb}.   

Resonances, or compound negative-ion states, play an important role in electron-molecule scattering \cite{schulz}.  They are identified as discrete metastable states of a molecule+electron that are embedded in, and interact with, a continuum of scattering states. Therefore, they have short lifetimes (in the range 10$^{-14}$s - 10$^{-12}$s) against decay by electron ejection, via different pathways, to a neutral state. Such decay often results in the neutral molecule being rovibrationally and/or electronically excited;  dissociative attachment is another possible exit channel. Resonances occur at a fixed collision energy and structures in elastic/inelastic differential cross sections (DCS) are their experimental signatures. In order to support discrete states in a continuum, an electron incident on a molecule should experience a sufficient attractive field. For neutral molecules, such interactions are rather weak. At low energies (0-3 eV), shape resonances occur when the incident electron is trapped in a potential barrier created by the combined effects of centrifugal, polarization, and exchange forces related to the ground electronic state of the molecule. Shape resonances can also be associated with electronically excited states. Rydberg states of molecules, on the other hand, support Feshbach resonances where the temporary negative ion arises from two electrons in Rydberg orbitals trapped in the field of the positive ion core. Feshbach resonances are normally bound by appreciable barriers ($\sim$200-500 meV) and occur at energies that lie just below the `parent' Rydberg state.  However, such trapping cannot be invoked for resonances near valence excited states. Resonances associated with these states have been postulated but several experimental attempts in the past to observe a Feshbach resonance below molecular valence states have yielded negative results \cite{schulz}. In our study we focus attention on the valence-excited $a^{3}\Pi$ state of CO, which lies 6.01 eV above the ground electronic state \cite {herzberg,eloss}. This state possesses a large electric dipole moment of 1.38 D and might, therefore, be considered a good candidate for temporarily binding an incoming electron in the interaction potential well. In contrast, the isoelectronic counterpart, the $B^{3}\Pi_{g}$ state in N$_{2}$ (7.35 eV), has no permanent electric dipole moment. 

24 years ago Wong and Schulz \cite{wong} studied the electron-impact DCSs of elastic and 
vibrational excitation ($\nu$=1,2) of CO in its ground electronic state ($^{1}\Sigma^{+}$) in the energy range 6-7 eV. They observed structures in the elastic and inelastic DCSs at energies that were coincident with the $\nu$=0,1 and 2 levels of the $a^{3}\Pi$ state of CO. These structures had the same cross section values at scattering angles $\theta$=50$^\circ$ and 90$^\circ$. Therefore, the opening of a channel resulting in excitation of the $a^{3}\Pi$ state was ruled out as the resulting p-wave character would have exhibited a strong decrease in the magnitude of the DCS at $\theta$=90$^\circ$. Wong and Schulz postulated that the structures in the DCSs arose from dipole-dominated resonances, and suggested that the potential-energy curve (PEC) of the negative ion must be coincident with that of the $a^{3}\Pi$ state, and must exhibit the same Franck-Condon factors. On the other hand, in the case of isoelectronic N$_2$ no sharp structures were observed in the elastic or vibrational excitation cross section functions in the vicinity of the corresponding valence $B^{3}\Pi_{g}$ state.  

We have employed an {\it ab initio} bound-state multireference configuration interaction (MRDCI) 
approach in our study. Out of various quartet states of different symmetries we identify the $^{4}\Sigma^{-}$ state of CO$^{-}$ which can be derived from the $^{3}\Pi$ state of CO. The electronic configuration of the latter state is (core)$^{4}(1\sigma)^{2}(2\sigma)^{2}(3\sigma)^{1}(1\pi)^{4}(2\pi)^{1}$. The lowest partially-filled antibonding orbital is $2\pi$ and there is a hole in the $3\sigma$ orbital. The resultant $^{4}\Sigma^{-}$ state of CO$^{-}$ is derived when an extra electron occupies the antibonding 2$\pi$ orbital. Consequently, one expects the equilibrium bond length (r$_{e}$) of the $^{4}\Sigma^{-}$ state to be longer than that of the neutral $a^{3}\Pi$ state. In order to compute the PECs for both the neutral and anion states we employed the correlation-consistent triple-$\zeta$ atomic basis set (CC-PVTZ) with  an augmented set of diffuse (spd) orbitals, that is, a [5s4p3d1f] set of contracted gaussian-type functions \cite{dunning} were placed on the two nuclei to obtain the molecular orbitals (MOs). The calculations were performed using the MRDCI package developed by Buenker, Peyerimhoff and coworkers \cite{buenker}. The technical details of our computations are as follows: We used 90 atomic basis functions to obtain the SCF MOs to generate the CI configuration space. For both CO and CO$^{-}$ the SCF MOs were optimized for the lowest $^{1}\Sigma^{+}$ and $^{2}\Pi$ configurations, respectively. The 1s core of both C and O was kept frozen and 4 high-lying MOs were excluded. Thus, for the CI step of the calculations, configurations were constructed from 84 MOs, effectively for 10 or 11 electrons. Depending on the distance and symmetry for the lowest 5 roots of a given irreducible representation of C$_{2v}$, a set of 30-65 relevant reference configurations was iteratively generated for the CI calculations. The threshold values applied for the selection were 10-30 $\mu$Hartree. The threshold energy values resulted from our predefined CI matrix dimension of 14000. We chose this value in order to keep the computational effort within reasonable limits while maintaining 
an acceptable, chosen accuracy of the final results. The final CI wave functions were well represented by the reference configuration space with squared coefficients of the main configurations being 0.91-0.94. The calculated ground-state CI value, including the MRDCI threshold extrapolation and the Davidson correction, at r$_{e}$=2.133 a$_{o}$, was 
-113.149427 au. The computed dissociation energy with respect to the zero-point energy of the ground-state was 11.015 eV, compared to the experimental value of 11.09 eV \cite{herzberg}. The r$_{e}$ value of the $^{3}\Pi$ state was 1.210 a$_{o}$ [r$_{e}$ (experimental)=1.206 a$_{o}$]. The potential energies of the neutral and anion states were calculated at intervals of 0.1 a$_{o}$ in the range 1.8-3.0 a$_{o}$, and a polynomial fit was employed to visualize the PECs.

It is important to note here the difficulties associated with computations of PECs of resonant molecular anion states \cite{list2}. With the addition of one electron to the neutral molecule, the amount of electron correlation increases, and computations must therefore involve larger basis sets with extended diffused functions. However, it has also been argued that the use of larger basis sets may not help in  describing  the actual anion state, but may actually describe a state of neutral molecule plus free electron (NMFE) \cite{mcweeny}. In other words, the computation may undergo a {\it variational collapse} in that the variational energy would tend towards the true ground state of the neutral molecule.  Such an effect has recently been pointed out in calculations of the PECs of H$_{2}^{-}$ \cite{mebel}. It may be that in order to achieve meaningful results a larger basis set should be employed but with a truncated limit 
for which a variational collapse can be avoided because of certain constraints (like orthogonality conditions, absence of continuum functions in the basis). We have found that on going from the CC-PVTZ basis to an augmented set with diffused sets of (sp) and (spd) functions, the $^{1}\Sigma^{+}$ and $a^{3}\Pi$ state of CO remained virtually unchanged while the $^{4}\Sigma^{-}$ state of CO$^{-}$ showed further stabilization with the use of more extended basis. However, we believe that with our augmented set of (spd) functions, the calculations are fairly stabilized. To ascertain this, we further augmented a diffuse set of f-functions, that is, computations were performed with AUG-CC-PVTZ basis set near the equilibrium distance of the 
$^{4}\Sigma^{-}$ state. These computations confirmed stabilization within the associated computed accuracy of a few tenths of an eV. A similar  AUG-CC-PVTZ basis has been recently employed to describe the $^{4}\Sigma_{g}^{-}$ and $^{4}\Pi_{u}$ states of N$_{2}^{-}$ \cite{list2}. There does not seem to be any variational collapse in that case and the computed state is not an NMFE state. To summarize, our study of CO$^-$ shows that (i) our computations are stabilized and (ii) the  $^{4}\Sigma^{-}$ state shows a distinctly larger r$_{e}$ value. It has been argued \cite{mebel} that a  free electron in the asymptotic limit would not significantly perturb r$_{e}$ of a neutral molecular state. We are therefore led to conclude that the present $^{4}\Sigma^{-}$ state describes a resonant CO$^{-}$ state and not a CO+free electron NMFE state. 

Our computed PECs along with associated vibrational energy levels are shown in Fig. 1. The energies of the $\nu$=0,1,2 and 3 levels of the $^{3}\Pi$ state of CO are 6.04, 6.26, 6.48, and 6.69 eV, respectively (with respect to the $\nu$=0 level of the ground electronic state). These energy levels match the sharp spikes observed in the DCSs measured by Wong and Schulz \cite{wong}. The computed Franck-Condon factors for these vibrational 
levels  with respect to $\nu$=0 of the ground state are 0.258, 0.339, 0.234, and 0.116, respectively. The FC factors for $\nu$=4 (0.042) and higher levels are very small. The  equilibrium distances that we compute for the $^{3}\Pi$ and $^{4}\Sigma^{-}$ state are 2.283 a$_{o}$ and 2.440 a$_{o}$, respectively. The energy difference between the minima ($\Delta E_{e}$) of these two states is 0.14 eV, and the difference between the zero-point energy levels ($\Delta E_{o}$) is 0.11 eV. The vibrational energy levels for $\nu$=0,1,2 and 3 levels of the $^{4}\Sigma^{-}$ state with respect to the $\nu$=0 of the ground state are at 6.15, 6.32, 6.49 and 6.66 eV, respectively. The energies of the $\nu$=1-3 levels are nearly coincident with those of the vibrational levels of the $^3\Pi$ state (fig. 1). Since none of these  states are accessible in the Franck-Condon region  we rationalize the occurrence of the Wong-Schulz resonance as a two step process: e+CO ($^{1}\Sigma^{+}$; $\nu$=0) $\rightarrow$ 
e + CO ($a^{3}\Pi$; $\nu$=0,1,2,3) $\rightarrow$ CO$^{-}(^{4}\Sigma^{-}; \nu =1,2,3)$. That is, on electron impact, the CO molecule is first excited into the accessible vibrational states of the lowest $a^{3}\Pi$ state, and then  capture of the electron  takes place into vibrational levels ($\nu$=0-3) of CO$^{-}$. Therefore, one would expect the resonances to occur at the vibrational energies of the $^4\Sigma^{-}$ state. Considering the experimental 
energy resolution of 0.055 eV for elastic scattering and 0.035 eV for vibrational excitation \cite{wong}, the location of the $\nu$=2 and 3 resonances match extremely well on the energy scale, but in the case of $\nu$=0 and 1, they are slightly displaced (by 0.1 eV and 0.03 eV, respectively). It is possible that the anion state may become further stabilized at even higher theoretical levels (employing more extended basis sets).  Nevertheless, the present computations establish the existence of a resonant $^{4}\Sigma^{-}$ state of CO$^{-}$. The isotropic nature of the measured DCSs can also now be rationalized: it results from the interplay between the resonant state of $\Sigma$ character and the $a^{3}\Pi$ state of neutral CO. 

We now consider recently observed signatures of long-lived (lifetime $>$few $\mu$s) CO$^{-}$ and N$_{2}^{~-}$ \cite{gnaser,vrb,middleton}. The extensive literature on e-CO (-N$_{2}$) collisions \cite{schulz} documents several resonances in these molecules, each possessing a lifetime of the order of a few femtoseconds. Recent observations of anions like N$_2^{~-}$ and CO$^-$ that are stable for at least a few microseconds suggest electronically excited precursor states.  Sommerfeld and Cederbaum  \cite{list2} have recently performed high-level computations for two possible quartet states of N$_{2}^{~-}$ ($^{4}\Pi_{u}$ and $^{4}\Sigma_{g}^{-}$) that are derived from the valence-excited states of N$_{2}$, $A^{3}\Sigma_{u}^{~+}$ and $B^{3}\Pi_{g}$, 
respectively.  One of the anion states ($^{4}\Pi_{u}$) was identified as a possible candidate for long lifetime, having a zero-point energy (ZPE) that was almost coincident with that of the neutral state. Calculations with further augmentation of basis indicated that this anion state 
might indeed possess a positive electron affinity. However, the computational accuracy was such that the sign of electron affinity could not be unambiguously determined. Sommerfeld and Cederbaum estimated the associated lifetime of the $^{4}\Pi_{u}$ state to be of the order of picoseconds. 

In the light of these observations one might expect that the $^{4}\Sigma^{-}$ state of CO$^{-}$ might also possess a positive electron affinity with respect to the $a^{3}\Pi$ parent state. The present study does point to this intriguing possibility, although much elaborate (and expensive) computations would be needed to settle the issue. If such lowering of the energy of the $^{4}\Sigma^{-}$ state does indeed occur, then the following mechanism might account for the longevity of the CO$^-$ anions in recent experiments. Collisional relaxation of the higher CO$^-$ levels to the ZPE level would become possible in the presence of a third body, $M$: CO$^{-}~(^{4}\Sigma^{-}; \nu >0)+ M\rightarrow CO^{-}(^{4}\Sigma^{-}; \nu=0)$. Such a possibility is, of course, remote in e-CO experiments but may be possible in the hot plasma environments that are created in the intense-field and sputtering experiments. Note that the Franck-Condon factors for the $^{4}\Sigma^{-}\rightarrow~X^1\Sigma^+$ transition are very unfavorable. 

Our results suggest several experimental and theoretical avenues that ought to be explored. Further work on metastable CO$^{-}$ states associated with higher excited states of CO is in progress in our laboratory.

\begin{figure}
\caption{Potential energy curves of the $X^1\Sigma^+$ and $a^{3}\Pi$ states of CO and of the $^{4}\Sigma^{-}$ state  of CO$^{-}$.}
\end{figure} 
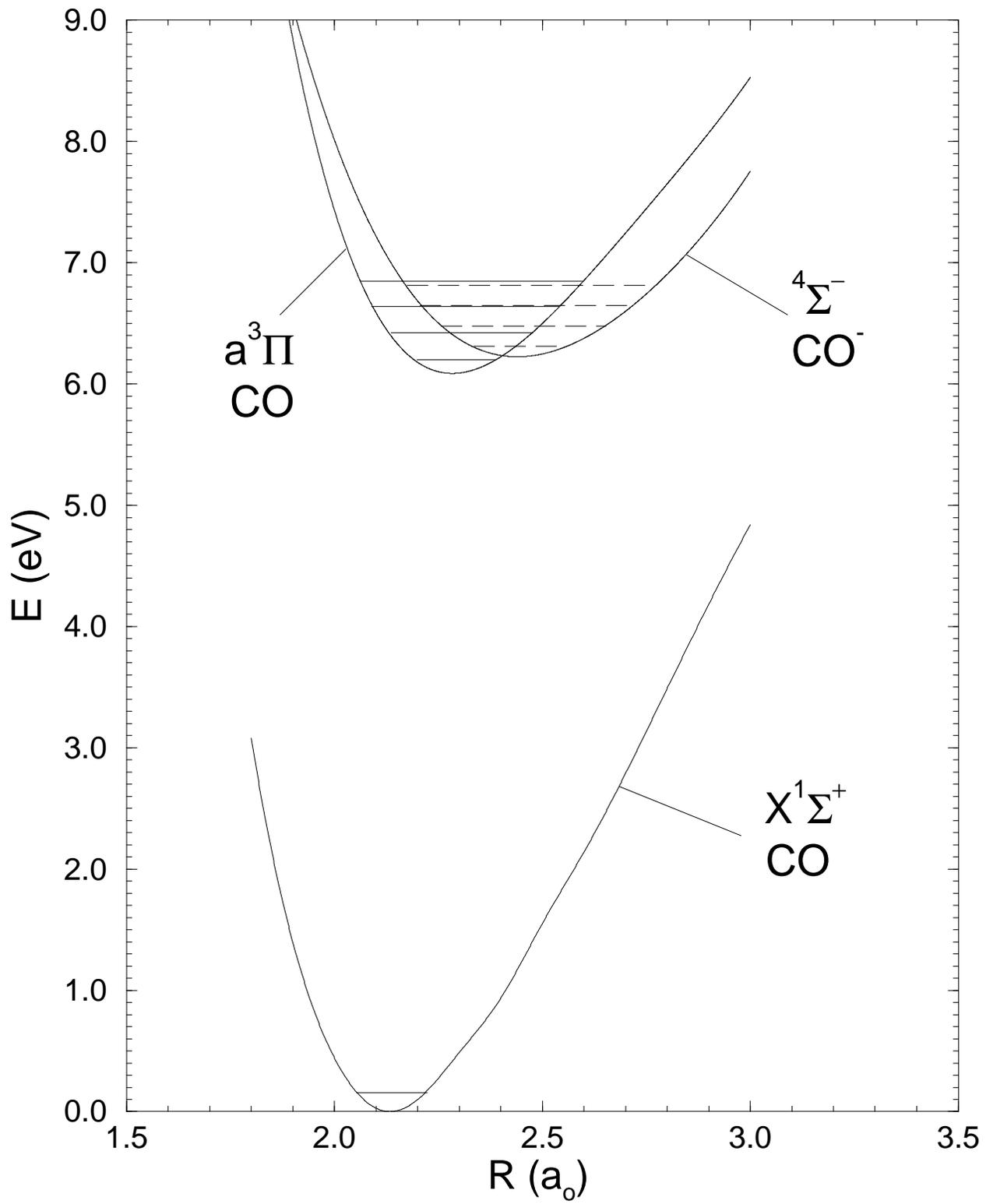
\end{document}